\documentclass{article}
\usepackage{arxiv}

\usepackage[numbers]{natbib}
\usepackage{filecontents}

\usepackage{tikz}
\usetikzlibrary{arrows,positioning,shadows,backgrounds,calc,shadings,shapes.arrows,shapes.symbols,fit}
\usetikzlibrary{shapes.geometric}
\usetikzlibrary{patterns}
\tikzset{
    >=stealth',
    punkt/.style={
           rectangle,
           rounded corners,
           draw=black, very thick,
           text width=6.5em,
           minimum height=2em,
           text centered},
    pil/.style={
           ->,
           thick,
           shorten <=2pt,
           shorten >=2pt,}
}
\usepackage{pgf-umlsd}

\usepackage{pgfplots}
\newlength\figureheight
\newlength\figurewidth
\usetikzlibrary{shapes}

\usepackage[nolist,nohyperlinks]{acronym}

\usepackage{array}
\usepackage{amssymb}
\usepackage{pifont}
\newcommand{\cmark}{\ding{51}}%
\newcommand{\xmark}{\ding{55}}%
\newcommand{\omark}{\ding{109}}%
\newcommand{\rmark}{\ding{71}}%

\usepackage{tabulary}
\newcolumntype{K}[1]{>{\arraybackslash}p{#1}}
\usepackage{courier}

\usepackage{graphicx}
\usepackage{float}
\usepackage{stfloats}    
\usepackage{placeins}
\usepackage[hidelinks]{hyperref}

\usepackage{url}
\makeatletter
\g@addto@macro{\UrlBreaks}{\UrlOrds}
\makeatother

\newcommand{\cm}{\textcolor{green}{\cmark}}
\newcommand{\xm}{\textcolor{red}{\xmark}}
\newcommand{\om}{\textcolor{black!50!white}{\omark}}
\newcommand{\tm}{\textcolor{orange}{\rmark}}

\usepackage{graphicx}

\newcommand\copyrighttext{%
  \footnotesize \textcopyright Efficient Passive ICS Device Discovery and Identification by MAC Address Correlation -- First published in the Electronic Workshops in Computing series at
  5th International Symposium for ICS \& SCADA Cyber Security Research 2018 (ICS-CSR 2018)
  DOI: \href{http://dx.doi.org/10.14236/ewic/ICS2018.3}{10.14236/ewic/ICS2018.3}
}
\newcommand\copyrightnotice{%
\begin{tikzpicture}[remember picture,overlay]
\node[anchor=south,yshift=10pt] at (current page.south) {\fbox{\parbox{\dimexpr\textwidth-\fboxsep-\fboxrule\relax}{\copyrighttext}}};
\end{tikzpicture}%
}

\title{Efficient Passive ICS Device Discovery and Identification by MAC Address Correlation}

\author{
Matthias Niedermaier \\
Matthias.Niedermaier@hs-augsburg.de \\
Hochschule Augsburg
\And
Thomas Hanka \\
Thomas.Hanka@hs-augsburg.de \\
Hochschule Augsburg
\And
Sven Plaga \\
Sven.Plaga@aisec.fraunhofer.de \\
Fraunhofer AISEC
\And
Alexander von Bodisco \\
Alexander.vonBodisco@hs-augsburg.de \\
Hochschule Augsburg
\And 
Dominik Merli \\
Dominik.Merli@hs-augsburg.de\\
Hochschule Augsburg
}

\begin{document}
\maketitle

\begin{abstract}
Owing to a growing number of attacks, the assessment of \acp{ICS} has gained in importance. 
An integral part of an assessment is the creation of a detailed inventory of all connected devices, enabling vulnerability evaluations. For this purpose, scans of networks are crucial.
Active scanning, which generates irregular traffic, is a method to get an overview of connected and active devices. Since such additional traffic may lead to an unexpected behavior of devices, active scanning methods should be avoided in critical infrastructure networks. 
In such cases, passive network monitoring offers an alternative, which is often used in conjunction with complex deep-packet inspection techniques. 
There are very few publications on lightweight passive scanning methodologies for industrial networks. 
In this paper, we propose a lightweight passive network monitoring technique using an efficient \ac{MAC} address-based identification of industrial devices. 
Based on an incomplete set of known \ac{MAC} address to device associations, the presented method can guess correct device and vendor information. 
Proving the feasibility of the method, an implementation is also introduced and evaluated regarding its efficiency. 
The feasibility of predicting a specific device/vendor combination is demonstrated by having similar devices in the database. 
In our \ac{ICS} testbed, we reached a host discovery rate of 100\% at an identification rate of more than 66\%, outperforming the results of existing tools.
\end{abstract}
\keywords{industrial control systems, vulnerability scanner, programmable logic controllers, security assessment}
\copyrightnotice
\acresetall

\section{Introduction}
\label{sec:introduction}
\begin{figure}[H]
  \centering
\begin{tikzpicture}[node distance=1cm,
    auto,
    block/.style={
      rectangle,
      draw=black,
      align=center,
      rounded corners,
      dashed
    },
    buffer/.style={
      draw,
      shape border rotate=90,
      isosceles triangle,
      isosceles triangle apex angle=20,
      fill=white!50!black,
      node distance=1cm,
      minimum height=4cm
    }
  ]
  
  \node[align=center, anchor=south] at (4,5.8) {ERP};
  \node[align=center, anchor=south] at (4,5.1) {MES};
  \node[align=center, anchor=south] at (4,4.4) {SCADA};
  \node[align=center, anchor=south] at (4,3.7) {PLC};
  \node[align=center, anchor=south] at (4,3) {\small Sensors, Actuators, etc.};
  \draw [-]  (2.0,2.9) -- (6.0,2.9);
  \draw [-]  (2.0,2.9) -- (4,7);
  \draw [-]  (6.0,2.9) -- (4,7);
  \draw [-]  (2.4,3.6) -- (5.6,3.6);
  \draw [-]  (2.8,4.3) -- (5.2,4.3);
  \draw [-]  (3.2,5.0) -- (4.8,5.0);
  \draw [-]  (3.6,5.7) -- (4.4,5.7);
   
  \node[align=center, anchor=south west,rotate=79] at (0.4,3.2) {time constraint};
  
  \coordinate (r0) at (0.3,2.9); 
  \coordinate (r1) at (1.0,6.3);
  \coordinate (r2) at (1.7,2.9);
  
  \filldraw[draw=white, shading = axis,rectangle, bottom color=black!35!white, top color=black!0!white] (r0) -- (r1) -- (r2) -- cycle;
  
  \node[align=center, anchor=south] at (1.0,3) {ms};
  \node[align=center, anchor=south] at (1.0,3.5) {s};
  \node[align=center, anchor=south] at (1.0,4.0) {m};
  \node[align=center, anchor=south] at (1.0,4.5) {h};
  \node[align=center, anchor=south] at (1.0,5.0) {d};
  
  \coordinate (t0) at (6.3,2.9); 
  \coordinate (t1) at (7.3,2.9);
  \coordinate (t2) at (7.3,6.3);
  
  \filldraw[draw=white, shading = axis,rectangle, bottom color=black!35!white, top color=black!0!white] (t0) -- (t1) -- (t2) -- cycle;  
  
  \node[align=center, anchor=south] at (7.1,3) {A};
  \node[align=center, anchor=south] at (7.1,3.8) {I};
  \node[align=center, anchor=south] at (7.1,4.6) {C};

  \coordinate (z0) at (6.3,2.9); 
  \coordinate (z1) at (6.3,6.3);
  \coordinate (z2) at (7.3,6.3);
  
  \filldraw[draw=white, shading = axis,rectangle, bottom color=black!0!white, top color=black!35!white] (z0) -- (z1) -- (z2) -- cycle;  

  \node[align=center, anchor=south] at (6.5,4.2) {A};
  \node[align=center, anchor=south] at (6.5,5) {I};
  \node[align=center, anchor=south] at (6.5,5.8) {C};
  
  \node[align=center, anchor=south west,rotate=90] at (6.3,4.7) {Office IT};
  \node[align=center, anchor=south west,rotate=90] at (7.8,3.0) {ICS};  
  
  \draw [->] (2.1,3.7) -- (3.7,7);
  \draw [->] (3.5,7) -- (1.9,3.7);
  \node[align=center, anchor=south west,rotate=64] at (2.4,4.6) {connected};

\end{tikzpicture}
\caption{IEC 62264 Industrial Automation Pyramid}
\label{fig_ot}
\end{figure}
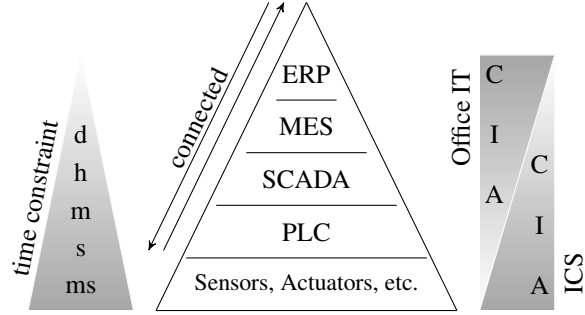

\ac{ICS} is the common umbrella term for the various devices used in industrial infrastructures. 
These generally consist of sensors, actuators, and embedded computers which are interconnected via network.
The basic hierarchical scheme is standardized by IEC 62264~\cite{international2003iec} 
and described as automation pyramid illustrated in \autoref{fig_ot}~\cite{Kellner:2009:VCE:1619258.1619271}. 
At the operational level of an industrial network, there are often \ac{ERP} and \ac{MES} layers.

Depending on the level of integration, \acp{ICS} also employs \ac{SCADA} or \acp{DCS}, \acp{PLC}, sensors, and actuators. 
Compared to other domains, the individual components of \acp{ICS} tend to have a long lifetime,
and cycles need to be processed in real time within certain time constraints.

Considering IT security protection goals, the established \ac{CIA} model is inverted~\cite{zhu2011taxonomy},
resulting Availability as the most important asset. 
In the past, most \ac{ICS} components were designed to operate in an isolated network.
Therefore, security was considered neglectable and development focused on functionality and on functional safety.
The increasing use of Internet-based technologies as baseline for the \ac{IIoT}, however, has enhanced the significance of security issues.
To observe the security situation of a network, security assessments are performed. 
These include different aspects, like situation analysis, risk identification and vulnerability scanning. 
The baseline for these actions is the discovery and identification of all devices present in the network. 
Technically, this is implemented either by active or passive scanning.

\textbf{Active scanning} broadcasts additional packets into a network environment and monitors the resulting traffic.
Depending on strict timing constraints, the injection of additional network traffic might lead to an unexpected behavior of the connected devices within industrial infrastructures. 
A standard network scan could lead to a \ac{DoS}, or result in defective devices or an incorrect behavior of the processes~\cite{Wedgbury:2015:AAD:2847563.2847571}.
As availability is considered the most important IT security protection goal, active scanning methods should be generally avoided~ \cite{niedermaier18woot}.

\textbf{Passive scanning} involves observing and capturing the packets transmitted within a certain period. 
This scanning scheme requires a device connected to the network to capture the network traffic. 
The capturing devices can be configured, limiting the captured traffic to certain packets of interest.
Compared to active scans, passive scanning schemes still have to deal with a larger amount of data. 
Hence, it requires a higher effort in extracting relevant information, thereby increasing processing time and the demands for computational power. 
Considering the IT security protection goal availability, these additional efforts are 
compensated as passive scanning schemes can also be integrated in fragile infrastructures without impairing regular communication. 

In this paper, we introduce an efficient and passive scanning and device identification scheme suitable for industrial Ethernet-based networks. 
On its baseline, it uses a \ac{MAC} address correlation method, enabling the discovery of \ac{ICS} components by analysis of their
\ac{ARP} broadcast packets using minimal configuration and integration effort. 
Without impairing the protection goal availability, the presented method is suitable for an integration to security assessment schemes of fragile or critical infrastructure networks.

The main \textbf{contributions} of the presented work are as follows:
\begin{itemize}
  \item We propose a passive scanning technique that can identify \ac{ICS} components without impairing the integrity of critical infrastructures.
  \item We show the feasibility of our method and provide a publicly available implementation of a \ac{PoC}.
  \item The used network captures from our testbed are made available to the public for further investigation.
\end{itemize}

The paper is organized in the following manner: 
State-of-the-art network scanners with a focus on passive fingerprinting are summarized in Section \ref{sec:relatedwork}. 
Section \ref{sec:backgroundknowledge} introduces the fundamentals of the \ac{MAC} addressing scheme which are crucial for further comprehension.
In Section \ref{sec:problem}, the challenges of \ac{MAC}-based device discovery and identification are introduced.
Section \ref{sec:methodology} presents the developed methodology. 
In Section \ref{sec:framework}, a \ac{PoC} implementation and testbed evaluations are presented. 
Finally, a conclusion and outlook is provided in Section \ref{sec:conclusion}. 

\section{Passive Network Scanning}
\label{sec:relatedwork}

This section provides a survey of currently available passive network scanners. Since the topic of device identification plays an important role, publicly accessible data sources are also listed. 

\textbf{NetworkMiner}~\cite{hjelmvik2008passive} is one of the most commonly used tools. Basically, it is an application-analyzing network to identify hosts. 
Using the combination of different fingerprinting methods and tools, NetworkMiner can determine the \ac{OS} that runs on a host, enabling vulnerability detection.
Since the protocol stack implementations are different for each \ac{OS}, the respective protocol header construction and length also differ.
The SYN/ACK packet-based identification takes advantage of different initial \ac{TTL} values for \ac{IP} and varying \ac{TCP} window sizes for \ac{TCP}. 
For these values, NetworkMiner uses the database of the \textbf{p0f}~\cite{p0f} tool.
Since different \acp{OS} also employ different implementations for \ac{DHCP}, identification is also possible to inspect these packets. Here, the device fingerprints from the \textbf{FingerBank}~\cite{Fingerbank} project are utilized.
A similiar approach to identifiy devices by their \ac{MAC} address is introcued by Martin et al.~\cite{martin2016decomposition}.

\textbf{SinFP}~\cite{Auffret2010} is a tool that supports active and passive \ac{OS} fingerprinting. The implementation of the two concepts increases the accuracy in situations when packets are altered by
mechanisms such as packet normalization or stateful packet inspection.
In these cases, SinFP sends several TCP probe frames to trigger different responses.
After the collection of all responses, the content of each packet is analyzed using an approach similar to p0f.
Additionally, SinFP supports a pure passive fingerprinting mode. In this configuration, only captured packets are analyzed which are obtained either from the network or a file.
Like the other tools, a database containing signatures and patterns is used to identify the detected devices.

After the capture of the input data for further analysis, different capture techniques are feasible.
A common approach involves using a mirror port provided by a switch or a network \ac{TAP}.
In an \ac{ICS} environment, however, port mirroring is not as effective as it would be in a conventional environment.
This effect is caused by a huge number of small groups of devices interconnected by simple switches~\cite{Wedgbury:2015:AAD:2847563.2847571}.

Compared to the investigated solutions, the proposed method has some major advantages.
First, a high identification rate of industrial components compared to existing tools is reached. 
Furthermore, an unused network port is sufficient for this scheme, because no special network switch feature or monitoring port is necessary. 
Moreover, no additional network traffic is generated in fragile \ac{ICS} networks. 
Finally, a low effort in database maintenance compared to deep packet inspection is sufficient.

\section{Media Access Control Addressing}
\label{sec:backgroundknowledge}

This section introduces the fundamentals of \ac{MAC} network device addressing, which is 
implemented in the data-link layer of the ISO/OSI reference model. 
These are utilized for device identification by the proposed method, which will be further described in the 
subsequent sections.

\subsection{\ac{MAC} Address Structure}
\begin{figure}[H]
  \centering
\begin{tikzpicture}[node distance=1cm,
    auto,
    block/.style={
      rectangle,
      draw=black,
      align=center,
      rounded corners,
      dashed
    }
  ]

\node[align=center, anchor=north west] at (2.1,7.5) {\small \texttt{Hexadecimal:}};
\node[align=left, anchor=north west] at (4.6,7.5) {\small \texttt{AC:DE:48:12:7B:80}};
\node[align=center, anchor=north west] at (2.1,7.1) {\small \texttt{Bit-reversed: }};
\node[align=left, anchor=north west] at (4.6,7.1) {\small \texttt{35:7B:12:48:DE:01}};
\node[text width=0.8cm, minimum width=1.2cm, align=center, anchor=north west] at (0.8,7.1) (a){Octet:};
\node[text width=0.8cm, minimum width=1.2cm, align=center, anchor=north west] at (0.8,6.6) (b){0};
\node[text width=0.8cm, minimum width=1.2cm, align=center, anchor=north west] at (2.0,6.6) (c){1};
\node[text width=0.8cm, minimum width=1.2cm, align=center, anchor=north west] at (3.2,6.6) (d){2};
\node[text width=0.8cm, minimum width=1.2cm, align=center, anchor=north west] at (4.4,6.6) (e){3};
\node[text width=0.8cm, minimum width=1.2cm, align=center, anchor=north west] at (5.6,6.6) (f){4};
\node[text width=0.8cm, minimum width=1.2cm, align=center, anchor=north west] at (6.8,6.6) (g){5};

\node[align=left, anchor=north west] at ([xshift=-0.05cm, yshift=-0.0cm]b.south west) (h){\small LSB};
\node[align=right, anchor=north east] at ([xshift=0.0cm, yshift=0.0cm]b.south east) (i){\tiny MSB};
\node[align=left, anchor=north west] at ([xshift=-0.0cm, yshift=0.0cm]c.south west) (j){\tiny LSB};
\node[align=right, anchor=north east] at ([xshift=0.0cm, yshift=0.0cm]c.south east) (k){\tiny MSB};
\node[align=left, anchor=north west] at ([xshift=-0.0cm, yshift=0.0cm]d.south west) (z){\tiny LSB};
\node[align=right, anchor=north east] at ([xshift=0.0cm, yshift=0.0cm]d.south east) (z){\tiny MSB};
\node[align=left, anchor=north west] at ([xshift=-0.0cm, yshift=0.0cm]e.south west) (z){\tiny LSB};
\node[align=right, anchor=north east] at ([xshift=0.0cm, yshift=0.0cm]e.south east) (z){\tiny MSB};
\node[align=left, anchor=north west] at ([xshift=-0.0cm, yshift=0.0cm]f.south west) (z){\tiny LSB};
\node[align=right, anchor=north east] at ([xshift=0.0cm, yshift=0.0cm]f.south east) (z){\tiny MSB};
\node[align=left, anchor=north west] at ([xshift=-0.0cm, yshift=0.0cm]g.south west) (z){\tiny LSB};
\node[align=right, anchor=north east] at ([xshift=0.05cm, yshift=0.0cm]g.south east) (z){\small MSB};

\draw [-,dashed] ([xshift=0.0cm, yshift=0.4cm]b.south east) -- ([xshift=0.0cm, yshift=-0.4cm]b.south east);
\draw [-,dashed] ([xshift=0.0cm, yshift=0.4cm]c.south east) -- ([xshift=0.0cm, yshift=-0.4cm]c.south east);
\draw [-,dashed] ([xshift=0.0cm, yshift=0.4cm]d.south east) -- ([xshift=0.0cm, yshift=-0.4cm]d.south east);
\draw [-,dashed] ([xshift=0.0cm, yshift=0.4cm]e.south east) -- ([xshift=0.0cm, yshift=-0.4cm]e.south east);
\draw [-,dashed] ([xshift=0.0cm, yshift=0.4cm]f.south east) -- ([xshift=0.0cm, yshift=-0.4cm]f.south east);

\node[align=center, anchor=north] at ([xshift=0.0cm, yshift=-0.5cm]b.south)
    (l){\scriptsize 00110101};
\node[align=center, anchor=north] at ([xshift=0.0cm, yshift=-0.5cm]c.south)
    (m){\scriptsize 01111011};
\node[align=center, anchor=north] at ([xshift=0.0cm, yshift=-0.5cm]d.south)
    (n){\scriptsize 00010010};
\node[align=center, anchor=north] at ([xshift=0.0cm, yshift=-0.5cm]e.south)
    (o){\scriptsize 01001000};
\node[align=center, anchor=north] at ([xshift=0.0cm, yshift=-0.5cm]f.south)
    (p){\scriptsize 11011110};
\node[align=center, anchor=north] at ([xshift=0.0cm, yshift=-0.5cm]g.south)
    (q){\scriptsize 00000001};

\draw [->] ([xshift=0.17cm, yshift=-0.8cm]l.south west) -- ([xshift=0.17cm, yshift=0.0cm]l.south west);
\draw [->] ([xshift=0.32cm, yshift=-0.4cm]l.south west) -- ([xshift=0.32cm, yshift=0.0cm]l.south west);    
\draw [-]  ([xshift=0.17cm, yshift=-0.8cm]l.south west) -- ([xshift=0.7cm, yshift=-0.8cm]l.south west); 
\draw [-]  ([xshift=0.32cm, yshift=-0.4cm]l.south west) -- ([xshift=0.7cm, yshift=-0.4cm]l.south west);  

\node[align=left, anchor=west] at ([xshift=0.7cm, yshift=-0.4cm]l.south west) (r){U/L address bit};
\node[align=left, anchor=west] at ([xshift=0.7cm, yshift=-0.8cm]l.south west) (s){I/G address bit};

\draw [<->] ([xshift=0.0cm, yshift=-2.0cm]b.south west) -- node[below]{MA-L: AC:DE:48} ([xshift=0.0cm, yshift=-2.0cm]d.south east);
\draw [-, dashed] ([xshift=0.0cm, yshift=-1.2cm]d.south east) -- ([xshift=0.0cm, yshift=-2.5cm]d.south east);

\draw [<->] ([xshift=0.0cm, yshift=-2.5cm]b.south west) -- node[below]{MA-M: AC:DE:48:1} ([xshift=0.0cm, yshift=-2.5cm]d.south east);
\draw [<->] ([xshift=0.6cm, yshift=-2.5cm]e.south west) -- ([xshift=0.0cm, yshift=-2.5cm]e.south east);
\draw [-, dashed] ([xshift=0.0cm, yshift=-1.2cm]e.south east) -- ([xshift=0.0cm, yshift=-3.0cm]e.south east);
\draw [-, dashed] ([xshift=0.6cm, yshift=-1.2cm]d.south east) -- ([xshift=0.6cm, yshift=-2.5cm]d.south east);

\draw [<->] ([xshift=0.0cm, yshift=-3.0cm]b.south west) -- node[below]{MA-S: AC:DE:48:12:7} ([xshift=0.0cm, yshift=-3.0cm]e.south east);
\draw [<->] ([xshift=0.6cm, yshift=-3.0cm]f.south west) -- ([xshift=0.0cm, yshift=-3.0cm]f.south east);
\draw [-, dashed] ([xshift=0.0cm, yshift=-1.2cm]f.south east) -- ([xshift=0.0cm, yshift=-3.0cm]f.south east);
\draw [-, dashed] ([xshift=0.6cm, yshift=-1.2cm]e.south east) -- ([xshift=0.6cm, yshift=-3.0cm]e.south east);

\end{tikzpicture}
\caption{Structure of an EUI-48~\cite{6847097}}
\label{fig_eui48}
\end{figure}
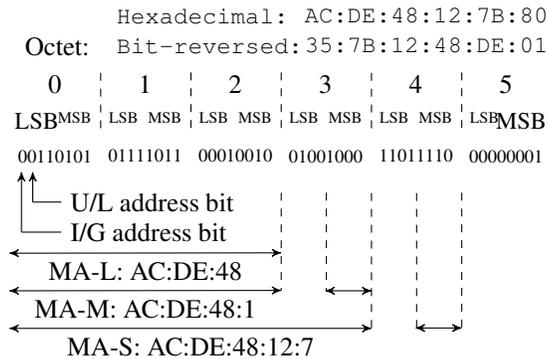

The \ac{MAC} address is a unique hardware address assigned to each network adapter. 
Nowadays, all known access methods with a \ac{MAC} layer (IEEE 802.1), such as Ethernet, Wi-Fi, and Bluetooth,
use the same MAC address format with a 48-bit MAC address as shown in \autoref{fig_eui48}.

The two least significant bits of the \ac{MAC} address determine the type of the address:
The \textit{Individual/Group address} \textbf{I/G bit} indicates whether the frame is transmitted as unicast (0) or multicast (1).
While a unicast frame will be sent to one specific device, a multicast frame is forwarded to a group of devices. 
An address consisting of 48 ones (FF:FF:FF:FF:FF:FF) is a broadcast address where all devices are addressed.
The \textit{Universally/Locally Administered} \textbf{U/L bit} is used to tell if the \ac{MAC} address
was taken from a fixed configuration (0) or dynamically chosen by the \ac{OS} (1).

The following address types can be observed when captured network traffic is examined.
If \textbf{I/G} is \textbf{0}, then it is an individual address (Unicast Address) for a network adapter. 
In contrast, if \textbf{I/G} is \textbf{1}, then the destination address is for a group of stations (Group address/Multicast address).
A universal, globally unique, and unchangeable \ac{MAC} address is used if \textbf{U/L}
is set to \textbf{0}, otherwise (\textbf{U/L} = \textbf{1}), it is locally changeable.

The assignment of \ac{MAC} address blocks is allocated and tracked by the \ac{IEEE} \ac{RA}. 
These are available in three different sizes, namely fixed length, MA-L (large), MA-M (medium), and MA-S (small), to meet the needs of vendors. 
The relation of the fixed  portion of a MAC address to the  corresponding number of possible MAC addresses is provided in \autoref{tab_macassign}.

\begin{table}[H]
\centering
\caption{IEEE \ac{MAC} Assignment~\cite{6847097}}
\label{tab_macassign}
\begin{tabular}{l c c c}
\hline \hline
\textbf{IEEE RA}    & \textbf{IEEE }         & \textbf{EUI-48}       & \textbf{Comp./organ.}    \\
\textbf{assignment} & \textbf{assigned}      & \textbf{block}        & \textbf{identifier}      \\
\hline
Company ID (CID)    & 24                     & 0                     & yes (CID)                \\[0.15cm]
Large (MA-L)        & 24                     & 2\textsuperscript{24} & yes (OUI)                \\[0.15cm]
Medium (MA-M)       & 28                     & 2\textsuperscript{20} & no                       \\[0.15cm]
Small (MA-S)        & 36                     & 2\textsuperscript{12} & yes (OUI-36 only)        \\
\end{tabular}
\end{table}

Information related to the current allocations, including the names of the respective vendors, can be obtained directly from \ac{IEEE} in different file formats~\cite{IEEE-All}. 
These files containing the currently allocated address blocks are associated with the contact information of their holders. 
This allows the mapping of an unknown \ac{MAC} address prefix of a network device to a manufacturer. 
Unfortunately, this approach does not allow any correlation with devices, since the vendor is allowed to freely assign addresses within the allocated space.

\subsection{Broadcast Messages}

In the data-link layer, broadcasting is the transmission of certain packets to all devices  
within a broadcast domain. A broadcast domain comprises hubs, switches, and bridges which can be divided by \acp{VLAN} or routers operating on the third layer. Broadcasting is used to accomplish different tasks of different protocols including:

\begin{itemize}
    \item \textbf{\ac{ARP}}, which is used if a network peer wants to communicate with another local device on a higher layer protocol. 
	To learn the \ac{MAC} address of the communication peer, an \ac{ARP}	request is sent to the Ethernet broadcast address which is forwarded by all network switches. 
	The results containing the respective \ac{MAC} addresses are cached in \ac{ARP} tables of the participating devices (RFC 826~\cite{plummer1982rfc}). 
Since these requests are essential for successful communication, all active \ac{MAC} addresses are cyclically propagated on the network.
    \item \textbf{\ac{DHCP}}, which allows the network configuration to be assigned to clients by a server (RFC 2131~\cite{droms1997rfc}).
    \item \textbf{Routing protocols}, which optimize the selection of proper routes for router communication.	
\end{itemize}

\section{Device Identification Challenges}
\label{sec:problem}
For the proposed MAC-based security assessment, two challenges were identified.
First, the approach has to determine devices from their \ac{MAC} addresses 
even if the vendor's assignment scheme remains unknown. 
Since passive scanning could be a time-consuming task, time estimation for a complete network device discovery was identified to be important.

\subsection{\ac{MAC} Address Vendor Assignment}
\begin{figure}[H]
  \centering
\begin{tikzpicture}[node distance=1cm,
    auto,
    block/.style={
      rectangle,
      draw=black,
      align=center,
      rounded corners,
      dashed
    }
  ]
  \node[align=center, anchor=west] at (0.2,4) (a){MA-L Vendor Assigned Part of a \ac{MAC} Address:};
  \node[align=center, anchor=west] at ([xshift=0.0cm, yshift=-0.3cm]a.south west) (b){\footnotesize XX:XX:XX:00:00:00};
  \node[align=center, anchor=east] at ([xshift=7.5cm, yshift=-0.3cm]a.south west) (c){\footnotesize XX:XX:XX:FF:FF:FF};
  \draw [->] ([xshift=0.1cm, yshift=-0.3cm]b.west) -- ([xshift=-0.1cm, yshift=-0.3cm]c.east);
  
  \draw [-, dashed] ([xshift=2.5cm, yshift=-0.2cm]b.south west) -- ([xshift=2.5cm, yshift=-1.8cm]b.south west);
  \draw [-, dashed] ([xshift=5.0cm, yshift=-0.2cm]b.south west) -- ([xshift=5.0cm, yshift=-1.8cm]b.south west);
  
  \node[text width=2.0cm, minimum width=2.5cm, align=center, anchor=north west] at ([xshift=0.0cm, yshift=-0.9cm]b.west) (d){Product-type X};
  \node[text width=2.0cm, minimum width=2.5cm, align=center, anchor=north west] at ([xshift=2.5cm, yshift=-0.9cm]b.west) (e){Product-type Y};
  \node[text width=2.0cm, minimum width=2.5cm, align=center, anchor=north west] at ([xshift=5.0cm, yshift=-0.9cm]b.west) (f){Product-type Z};
  
  \draw [->] ([xshift=0.1cm, yshift=0cm]d.north west) -- ([xshift=0.0cm, yshift=0cm]d.north east);
  \node[align=center, anchor=south west] at ([xshift=-0.00cm, yshift=0.01cm]d.north west) {\scriptsize 00:00:00};
  \node[align=center, anchor=south east] at ([xshift=0.05cm, yshift=0.01cm]d.north east) {\scriptsize 5F:FF:FF};
  
  \draw [->] ([xshift=0.0cm, yshift=0cm]e.north west) -- ([xshift=0.0cm, yshift=0cm]e.north east);
  \node[align=center, anchor=south west] at ([xshift=-0.05cm, yshift=0.01cm]e.north west) {\scriptsize 60:00:00};
  \node[align=center, anchor=south east] at ([xshift=0.05cm, yshift=0.01cm]e.north east) {\scriptsize BF:FF:FF};
  
    \draw [->] ([xshift=0.0cm, yshift=0cm]f.north west) -- ([xshift=-0.1cm, yshift=0cm]f.north east);
  \node[align=center, anchor=south west] at ([xshift=-0.05cm, yshift=0.01cm]f.north west) {\scriptsize C0:00:00};
  \node[align=center, anchor=south east] at ([xshift=0.0cm, yshift=0.01cm]f.north east) {\scriptsize FF:FF:FF};
  
  \end{tikzpicture}
\caption{\ac{MAC} Address Vendor Assignment}
\label{fig_macassignment}
\end{figure}
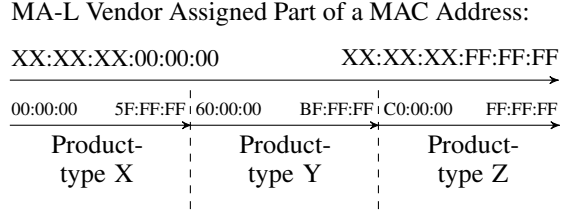

Each of the devices is programmed with a unique \ac{MAC} address taken from the \ac{IEEE}-assigned address blocks.
Parts of the available blocks are often used sequentially for various products from a company.
For example, regarding MA-L ranges, \autoref{fig_macassignment} illustrates such a possible block-wise assignment of different product types.
Product types X, Y, and Z are separated in a sequential and continuous block, which are often spread across the full range.

Since this information is unknown, \ac{MAC} addresses have to be collected. 
A correlation of yet unknown devices has to be determined from that source either by direct lookup or 
a proper approximation scheme. 

\subsection{Interarrival Time of Packets}
\label{sec:interarrival}
Estimating the time for complete device discovery coverage, the interarrival time $t_{arrival}$ was found to be an important key figure.
  
 \begin{eqnarray} \label{eq:arrival_1}
    t_{arrival} = t_{packet_{n}} - t_{packet_{n-1}} 
 \end{eqnarray}
 
Generally, $t_{arrival}$ is defined as the elapsed time between the arrival of two consecutive packets containing the same information. 
It is calculated in the manner shown by \autoref{eq:arrival_1}. 
The mean interarrival time $\overline{t}_{arrival}$ is composed of constituent measurements ($measure$) shown by \autoref{eq:arrival_2}.

\begin{eqnarray} \label{eq:arrival_2}
    \overline{t}_{arrival} =  \frac{1}{n} \cdot \sum_{n=1}^{n=measure} t_{interarrival_{n}}
\end{eqnarray}

\autoref{eq:arrival_3} calculates the maximum time to get a packet of one device.

\begin{equation} \label{eq:arrival_3}
   t_{arrival_{max}} = \max_{\forall measure \in device} t_{arrival}(measure)
\end{equation}

Finally, the expected time for a complete network device discovery  $t_{coverage}$ is calculated (see \autoref{eq:arrival_4}) by using the maximum interarrival time of all devices within the monitored network segment.

\begin{equation} \label{eq:arrival_4}
   t_{coverage} = \max_{\forall device \in network} t_{arrival_{max}}(device)
\end{equation}

\begin{table*}[tb]
\centering
\caption{Devices Employed within the Testbed}
\label{tab_racksetup}
\resizebox{\textwidth}{!}{%
\begin{tabular}{l l | >{\ttfamily}c >{\ttfamily}c >{\ttfamily}c | >{\ttfamily}c >{\ttfamily}c >{\ttfamily}c | >{\ttfamily}c}
\hline \hline
\textbf{Vendor}           & \textbf{Product}        & \textbf{No.} & \textbf{MAC}      & \textbf{IP}          & \textbf{No.} & \textbf{MAC}      & \textbf{IP}    & \textbf{Distance}  \\
                          &                         &              & \multicolumn{2}{c|}{Rack 1} &             & \multicolumn{2}{c|}{Rack 2}  & hex                \\                        
\hline  
Siemens                   & CPU 1211C               & 00           & 28:63:36:C6:C7:D4 & 192.168.0.10         & 12           & 28:63:36:C6:CC:67 & 192.168.0.110  & 0x000493             \\ 
Siemens                   & KP 300                  & 01           & 00:1C:06:35:C0:7C & 192.168.0.11         & 13           & 00:1C:06:35:C0:7B & 192.168.0.111  & 0x000001             \\
Phoenix                   & ILC 151                 & 02           & 00:A0:45:9D:40:74 & 192.168.0.20         & 14           & 00:A0:45:9D:42:54 & 192.168.0.120  & 0x0001E0             \\
ABB                       & PM554-T                 & 03           & 00:24:59:0A:4C:B7 & 192.168.0.21         & 15           & 00:24:59:0A:58:B0 & 192.168.0.121  & 0x000BF9             \\
Crouzet                   & em4 B26-2GS             & 04           & 84:AC:FB:00:05:E0 & 192.168.0.22         & 16           & 84:AC:FB:00:05:E6 & 192.168.0.122  & 0x000006             \\
Siemens                   & LOGO! 24RCE             & 05           & E0:DC:A0:1C:35:85 & 192.168.0.23         & 17           & E0:DC:A0:1C:35:4F & 192.168.0.123  & 0x000036             \\
Wago                      & 750-889                 & 06           & 00:30:DE:0C:AA:68 & 192.168.0.30         & 18           & 00:30:DE:0C:AA:6C & 192.168.0.130  & 0x000004             \\
Wago                      & 750-8100                & 07           & 00:30:DE:41:B9:F0 & 192.168.0.31         & 19           & 00:30:DE:41:B9:E6 & 192.168.0.131  & 0x00000A             \\
Wago                      & 750-880                 & 08           & 00:30:DE:0C:AE:84 & 192.168.0.32         & 20           & 00:30:DE:0C:AE:68 & 192.168.0.132  & 0x00001C             \\
Schneider                 & TM221CE16T              & 09           & 00:80:F4:0E:58:89 & 192.168.0.50         & 21           & 00:80:F4:0E:59:BC & 192.168.0.150  & 0x000133             \\
Schneider                 & HMISTU855               & 10           & 00:01:23:2D:BA:A3 & 192.168.0.51         & 22           & 00:01:23:2D:BD:7B & 192.168.0.151  & 0x0002D8             \\
Moxa                      & NP5110                  & 11           & 00:90:E8:2A:E5:34 & 192.168.0.70         & 23           & 00:90:E8:56:78:5D & 192.168.0.170  & 0x2B9329             \\
\end{tabular}
} 
\end{table*}
\normalsize

\section{Using MAC Addresses for Device Discovery and Identification}
\label{sec:methodology}
This section introduces the methodology developed for the \ac{MAC} address-based security assessments. 
Essentially, it builds up on capturing \ac{MAC} broadcasts. 
The \ac{MAC} addresses are extracted from these captures and fed to an address-identification process. 
These are mapped to known vulnerabilities of the identified devices.

\subsection{Passive Scan Utilizing ARP Broadcast Messages}
In an industrial network, there are mostly static routes with fixed network configurations.
Consequently, \ac{DHCP} and routing protocol broadcasts are rare.
In contrast, \ac{ARP} requests are performed in cases where no \ac{MAC} address 
is cached for a certain device. The \ac{ARP} cache contains a four-column table containing 
the protocol type, the protocol address of the sender, the hardware address of the sender, and the entry time. 
The expiration time for the entries is not specified by the relevant
RFC 826~\cite{plummer1982rfc}. 

\subsection{\ac{MAC} Address Correlation}
\label{sec:methodology_correlation}
To successfully correlate unknown \ac{MAC} addresses to devices, a prebuilt data-pool of known devices is essential.

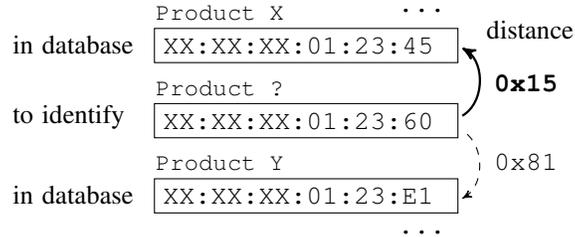
\begin{figure}[H]
  \centering
\begin{tikzpicture}[node distance=1cm,
    auto,
    block/.style={
      rectangle,
      draw=black,
      align=center,
      rounded corners,
      dashed
    }
  ]
  \node[anchor=south east] at (5, 2.55) {\texttt{...}};
  \node[draw,text width=3.8cm, anchor=south west] at (1, 2) {\texttt{XX:XX:XX:01:23:45}};
  \node[text width=3.8cm, anchor=south west] at (-1, 2) {in database};
  \node[anchor=south west] at (0.9, 2.45) {\texttt{\small Product X}};
  
  \node[draw,text width=3.8cm, anchor=south west] at (1, 1) {\texttt{XX:XX:XX:01:23:60}};
  \node[text width=3.8cm, anchor=south west] at (-1, 1) {to identify};
  \node[anchor=south west] at (0.9, 1.45) {\texttt{\small Product ?}};
  
  \node[draw,text width=3.8cm, anchor=south west] at (1, 0) {\texttt{XX:XX:XX:01:23:E1}};
  \node[text width=3.8cm, anchor=south west] at (-1, 0) {in database};
  \node[anchor=south west] at (0.9, 0.45) {\texttt{\small Product Y}};
  \node[anchor=south east] at (5, -0.4) {\texttt{...}};
  
  \draw [<-, bend angle=60, dashed, bend right]  (5.1,0.2) to (5.1,1.1);
  \draw [<-, bend angle=60, line width=0.3mm,bend left]  (5.1,2.2) to (5.1,1.3);
  
  \node[anchor=south west] at (5.3, 2.2) {distance};
  \node[anchor=south west] at (5.4, 0.45) {\texttt{0x81}};
  \node[anchor=south west] at (5.4, 1.5) {\texttt{\textbf{0x15}}};
  \end{tikzpicture}
\caption{\ac{MAC} Based Identification}
\label{fig_distance}
\end{figure}

\autoref{fig_distance} illustrates the developed device identification scheme. 
In this example, there are two initially known \ac{MAC} addresses stored in the data-pool.
Both addresses are close to a \ac{MAC} address of an unknown device. 
The illustrated distance value is calculated from the difference between the addresses. 
A smaller distance increases the chance of a correct device identification. 
In the present case, the unidentified device is more likely to be the same product
family as ``Product X.'' This correlation is used for the passive identification of \ac{ICS} components.

\subsection{Mapping of Vulnerabilities}
\begin{figure}[H]
  \centering
\begin{tikzpicture}[node distance=1cm]  
  \node[draw,circle,minimum size=1.4cm,inner sep=0pt, align=center] at (0,0) (a){\small Vendor \\ \small and product \\ \small name};
  \node[draw,circle,minimum size=1.4cm,inner sep=0pt, align=center] at (2,0) (b){\small CPE};
  \node[draw,circle,minimum size=1.4cm,inner sep=0pt, align=center] at (4,0) (c){\small CVE};
   
  \draw [->, bend angle=60, dashed, bend left]  (a) to (b);
  \node[anchor=south] at (2.1, 0.8) {search};
  \draw [->, bend angle=60, dashed, bend right]  (b) to (c);
  \node[anchor=north] at (4, -0.8) {relation};
  \end{tikzpicture}
\caption{Method of Vulnerability Mapping}
\label{fig_cpecve}
\end{figure}
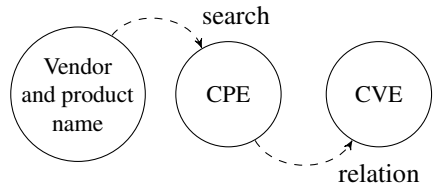

With the determined device correlation, the identification of known vulnerabilities is feasible.
\autoref{fig_cpecve} shows the relation of the vulnerability mapping.
With information concerning the vendor and product name, the corresponding \ac{CPE} is searched using pattern recognition.
The relationship between \ac{CPE} and \ac{CVE} allows the assignment of vulnerabilities to a device.

\section{Framework and Evaluation}
\label{sec:framework}
To evaluate the feasibility of the presented method, the outlined methodology is 
implemented in a framework.  
After determining its efficiency, an industrial testbed was created where the framework was further evaluated.

\subsection{Testbed Setup / Preliminary Investigations }

To facilitate an examination of the \ac{MAC} address distribution, the testbed contains two racks, each of which has an identical set of 12 devices. 
The components of the rack-mounted testbed are listed in \autoref{tab_racksetup}~\cite{niedermaier2018cort}.
The last column denotes the distance, which is the absolute value calculated from the
subtraction of the \ac{MAC} addresses of devices of the same type.
Except for those from Moxa, the devices were bought at the same time.
Therefore, the distances between the \ac{MAC} addresses are mostly small.
Further inspection of the Moxa devices revealed different production dates and
higher values for the distance.

\subsection{Interarrival Time of \ac{ARP} Packets}
\label{sec:interarrivalarp}

First, $t_{arrival}$ between consecutive \ac{ARP} broadcasts for each of the 
connected devices is determined. 
For this task, \ac{MAC} addresses are extracted from the \ac{ARP} broadcasts captured from a 12-hour \ac{pcap}.

\begin{figure*}[tb]
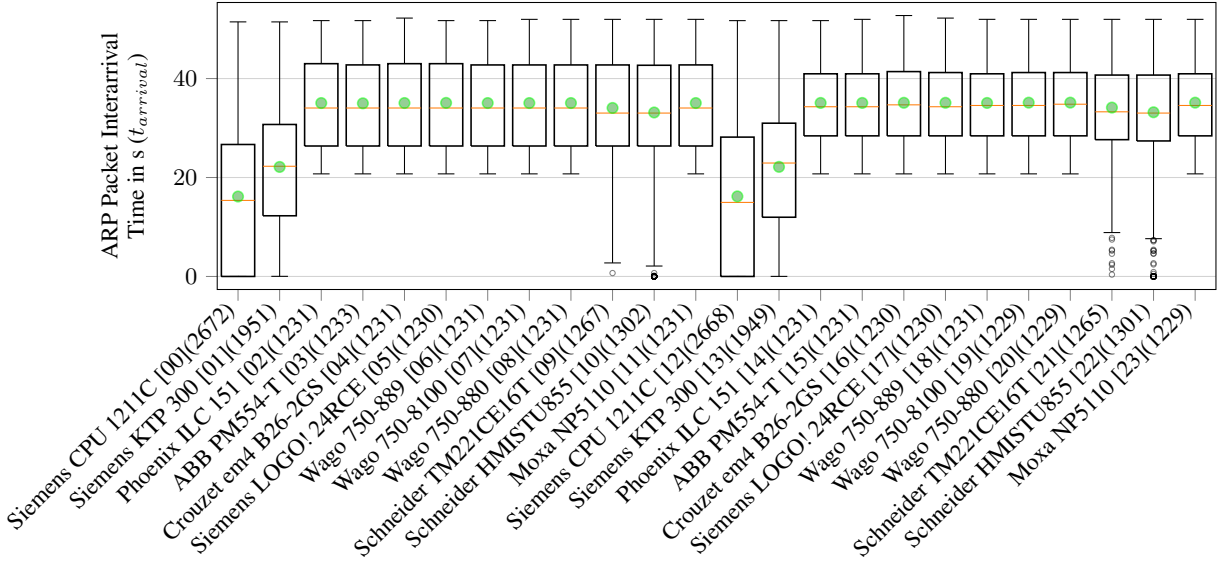

  \centering
\begin{small}
\setlength\figureheight{5.4cm}
\setlength\figurewidth{0.90\textwidth}
\InputIfFileExists{boxplot_arp_big.tikz}{}{\textbf{!! Missing graphics !!}}
\end{small}
\caption{ARP Packet Occurrence by Device Over Time [Device Number](Total ARP Packets)}
\label{fig_arpboxplot_big}
\end{figure*}

The results of this experiment consist of 3559310 packets, where 220988 (6.21\%) are \ac{ARP} packets.
\autoref{fig_arpboxplot_big} illustrates the calculated $t_{arrival}$ for the 24 devices.
In 70\% of the cases, an \ac{ARP} packet shows up every 16--37 seconds (${\overline{t}_{arrival}}$), not exceeding the boundary of one minute ($t_{arrival_{max}}$) at the worst cases.
With this data, $t_{coverage}$ of this network is about one minute.
The total number of \ac{ARP} packets varies from about 1200 to 2700 over the measurement time of 12 hours.
In the testbed, a continuous \ac{SCADA} monitoring process is implemented, which queries the status of all devices every second. 
This communication generates constant traffic, which could lead to a homogeneous plot.
There are four \acp{HMI}, the Siemens KP 300 [01,13] and the Schneider HMISTU855 [10,22], which communicate with the controllers Siemens CPU 1211C [00,12] and Schneider TM22ICE16T [09,22].
Because of the constant communication between them, there are more \ac{ARP} packets of these devices in the \ac{pcap}.

\subsection{\ac{MAC} Address Database}
\label{sec:MAC_Database}
\begin{figure*}[tb]
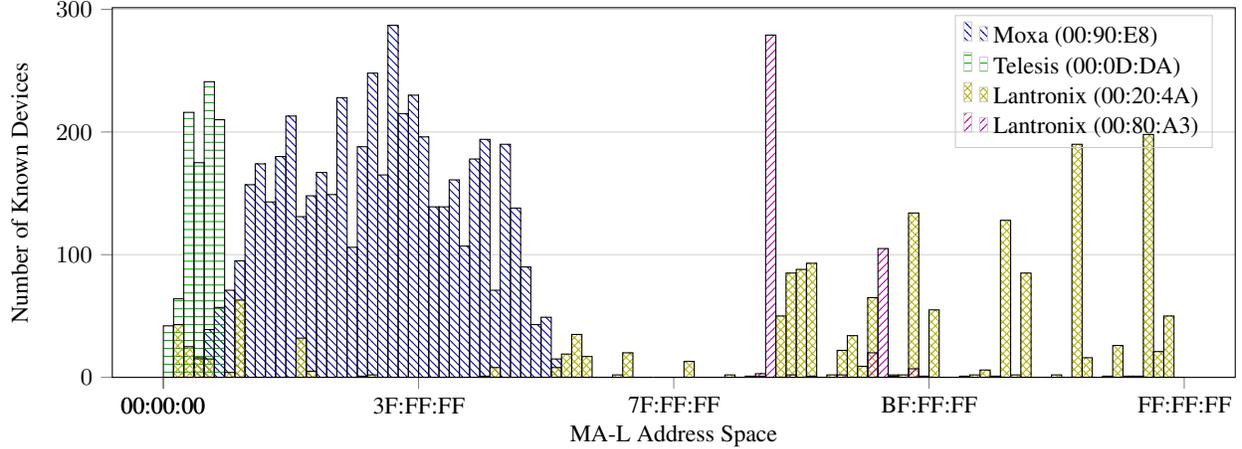

  \centering
\begin{small}
\setlength\figureheight{6.5cm}
\setlength\figurewidth{\textwidth}
\InputIfFileExists{macplot.tikz}{}{\textbf{!! Missing graphics !!}}
\end{small}
\caption{Histogram of Known Devices \ac{MAC} Addresses in Their MA-L Address Space}
\label{fig_macDistribution}
\end{figure*}

The initial device database (see \autoref{tab_database}) was built using data from Censys~\cite{censys15}, Shodan~\cite{matherly2009shodan}, Google image and marketplace search, as well as previous scans from our research group. 
This database is locally stored and the baseline for the identification methodology is provided in Section \ref{sec:methodology_correlation}.
\autoref{fig_macDistribution} illustrates the distribution of known devices in the local database over the complete MA-L range of the corresponding vendor, with more than 500 entries.
Some vendors have more than one MA-L range, resulting in more company identifiers.

\begin{table}[H]
\centering
\caption{Number of Devices and Known Vulnerabilities in the Database}
\label{tab_database}
\begin{tabular}{l l r r}
\hline \hline
\textbf{Vendor}           & \textbf{Products}        & \textbf{Devices} & \textbf{Known Vuln.}  \\
\hline
MOXA Technologies         & Networking equipment     & 5242             &  80             \\
Lantronix                 & Networking equipment     & 2126             &   6             \\
Allied Telesis            & Networking equipment     & 950              &   9             \\
Siemens                   & Automation equipment     & 75               & 242             \\
WAGO Kontakttechnik       & Automation equipment     & 49               &   5             \\ 
Star Micronics            & (Receipt) printer        & 20               &   1             \\ 
Schneider Electric        & Automation equipment     & 14               & 113             \\
Phoenix Contacts          & Automation equipment     & 11               &  6              \\
ABB                       & Automation equipment     & 6                & 18              \\
Hirschmann Industries     & Networking equipment     & 3                &   8             \\
Crouzet                   & Automation equipment     & 3                &   0             \\
\hline 
                          & \textbf{Total:}          & \textbf{8499}    & \textbf{488}    \\
\end{tabular}
\end{table}

The dataset plots of Moxa and Allied Telesis are bundled across the lower \ac{MAC} address range of the complete MA-L space.
This indicates that the examined manufactures have yet not exceeded their assigned address space.
Moreover, the available database entries suggest a linear assignment process. 

The results for Lantronix devices indicate a different assignment process.
The \ac{MAC} addresses are spread over the full MA-L space, while the majority of addresses are concentrated at the upper range.
This indicates a kind of randomization, as there are no larger connected groups of addresses. 
However, the entries in our dataset show a possible sequential block-wise assignment, which could also be a sign that our dataset is not large enough to include many devices of the same range.
Eventually, it is not possible to make a final statement on the MAC distribution policy applied by Lantronics.

\subsection{\acl{PoC} Framework}
\begin{figure}[H]
  \centering
  \small
\begin{tikzpicture}[node distance=1cm,
    auto,
    block/.style={
      rectangle,
      draw=black,
      align=center,
      rounded corners,
      dashed
    }
  ]
    \node[block, draw, align=center, minimum width=1.5cm, minimum height=0.5cm, anchor=south west] at (-2.0, 1.1)   (a) {Pcap-ng};
    \node[block, draw, align=center, minimum width=2.0cm, minimum height=0.5cm, anchor=south west] at (-2.5, 0.5)   (b) {Live capture};
    \node[block, draw, align=center, minimum width=4cm, minimum height=0.5cm, anchor=south west] at (0, 0.8)   (c) {Analyze capture};
	\draw [->] (a.east) -- ([xshift=0.0cm, yshift=0.05cm]c.west);
	\draw [->] (b.east) -- ([xshift=0.0cm, yshift=-0.05cm]c.west);
	
	\node[block, draw, align=center, minimum width=1.5cm, minimum height=0.5cm, anchor=south west] at (-2.0, 1.8)   (d) {IEEE};
	\node[block, draw, align=center, minimum width=4cm, minimum height=0.5cm, anchor=south west] at (0, 1.8)   (e) {Fetch and parse MAC lists};
	\draw [->] (d.east) -- ([xshift=0.0cm, yshift=-0.00cm]e.west);
	
	\node[block, draw, align=center, minimum width=2.0cm, minimum height=0.5cm, anchor=south west] at (-2.5, 2.95)   (f) {cve-search};
	\node[block, draw, align=center, minimum width=4cm, minimum height=1.0cm, anchor=south west] at (0, 2.7)   (g) {Fetch and parse \\ vulnerability list};
    \draw [->] (f.east) -- ([xshift=0.0cm, yshift=-0.00cm]g.west);	
	
	\node[block, draw, align=center, minimum width=2.0cm, minimum height=0.5cm, anchor=south west] at (-2.5, 4.0)   (h) {local database};
	\node[block, draw, align=center, minimum width=4cm, minimum height=0.5cm, anchor=south west] at (0, 4.0)   (i) {Parse device list};
	\draw [->] (h.east) -- ([xshift=0.0cm, yshift=-0.00cm]i.west);
	
	\node[block, draw, align=center, minimum width=6.5cm, minimum height=0.5cm, anchor=south west] at (-2.5, -0.1)   (j) {Merge data and get device identification};
	
	\node[block, draw, align=center, minimum width=6.5cm, minimum height=0.5cm, anchor=south west] at (-2.5, -0.9)   (k) {Map vulnerabilities to device};

	\draw [->] (i.south) -- (g.north);
	\draw [->] (g.south) -- (e.north);
	\draw [->] (e.south) -- (c.north);
	\draw [->] (c.south) -- (j.north);
	\draw [->] (j.south) -- (k.north);
\end{tikzpicture}
\caption{Dataflow within the Framework}
\label{fig_highlvl}
\normalsize
\end{figure}
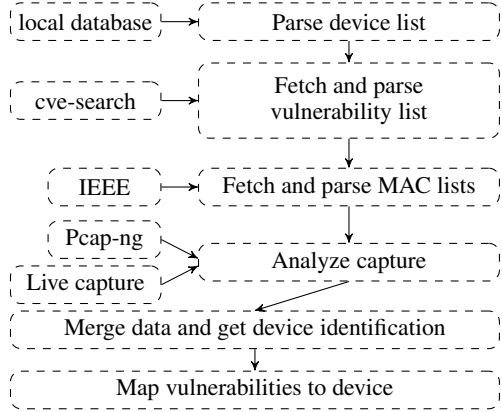

The \acf{PoC} framework is written in Python 3 and implements automatic evaluation of \ac{MAC}-based discovery and identification of \acp{ICS}.
It takes a \ac{pcap} file or input, or captures live traffic, analyzes it, and creates a security report on the detected network devices.
The pseudo program structure of the framework is shown in \autoref{fig_highlvl}.

At startup, the local \ac{MAC} databases (see Section \ref{sec:MAC_Database}) are imported. 
These are extensible \ac{CSV} files that store devices with their product name and \ac{MAC} address.
Current vulnerabilities are fetched automatically from cve details~\cite{cve}.
In the next step, the \ac{IEEE} \ac{MAC} vendor lists, which are publicly available on the Internet, are downloaded and parsed.
Subsequently, logged or live network traffic is analyzed using the pcapy tool~\cite{pcapy}, which provides access to the \ac{pcap} packet capture library. 

The \ac{MAC} addresses of the devices found in the capture are now compared with the local database of known devices. 
If a \ac{MAC} address of a device from the network capture is at a predefined distance from a \ac{MAC} address of a known device, it is possible that these devices are the same product. 
Based on this information, the \ac{CVE} database is used to map possible vulnerabilities to the devices.

\subsection{Network Integration}
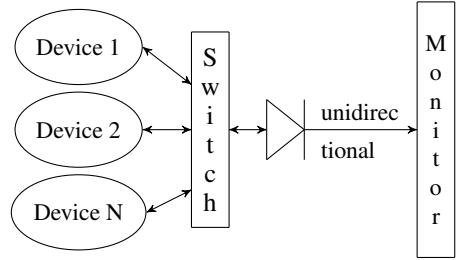
\begin{figure}[H]
  \centering
\begin{tikzpicture}[node distance=1cm,
    auto,
    block/.style={
      rectangle,
      draw=black,
      align=center,
      rounded corners,
      dashed
    }
  ]
  \node[draw,ellipse,minimum size=1.0cm,inner sep=1pt, align=center] at (1,2.5) (a){\small Device 1};
  \node[draw,ellipse,minimum size=1.0cm,inner sep=1pt, align=center] at (1,1.4) (b){\small Device 2};
  \node[draw,ellipse,minimum size=1.0cm,inner sep=1pt, align=center] at (1,0.3) (c){\small Device N};
  
  \draw [<->] ([xshift=0.0cm, yshift=0.0cm]a.east) -- (2.5,2.0);
  \draw [<->] ([xshift=0.0cm, yshift=0.0cm]b.east) -- (2.5,1.4);
  \draw [<->] ([xshift=0.0cm, yshift=0.0cm]c.east) -- (2.5,0.6);
  
  \draw [-] (2.5,0.1) -- (2.5,2.7);
  \draw [-] (2.5,0.1) -- (3.0,0.1);
  \draw [-] (3.0,0.1) -- (3.0,2.7);
  \draw [-] (2.5,2.7) -- (3.0,2.7);
  \node[align=center, anchor=center] at (2.75,1.4) { S \\ w \\ i \\ t \\ c \\ h};
  
  \draw [-] (3.5,1.0) -- (3.5,1.8);
  \draw [-] (3.5,1.0) -- (4.0,1.4);
  \draw [-] (3.5,1.8) -- (4.0,1.4);
  \draw [-] (4.0,1.0) -- (4.0,1.8);
  
  \draw [<->] (3.0,1.4) -- (3.5,1.4);
  \draw [->] (4.0,1.4) -- (5.5,1.4);
  \node[align=center, anchor=south west] at (4.1,1.4) {\footnotesize unidirec};
  \node[align=center, anchor=south west] at (4.1,0.9) {\footnotesize tional};

  \draw [-] (5.5,-0.3) -- (5.5,3.1);
  \draw [-] (5.5,-0.3) -- (6.0,-0.3);
  \draw [-] (6.0,-0.3) -- (6.0,3.1);
  \draw [-] (5.5,3.1)  -- (6.0,3.1);
  \node[align=center, anchor=center] at (5.75,1.4) {\small M \\ \small o \\ \small n \\ \small i \\ \small t \\ \small o \\ \small r};
  \end{tikzpicture}
\caption{Passive Network Monitoring}
\label{fig_networkint}
\end{figure}

The network integration for this method is easier than most of the available passive network monitoring tools, because no mirror port or network \ac{TAP} is necessary.
Basically, all broadcast messages containing \ac{MAC} addresses are sufficient for device discovery. 
Figure \ref{fig_networkint} shows the integration possibility of the proposed framework in an existing network.

To ensure a feedback-free passive scan, a unidirectional network diode can be placed between the switch and the scan device. 
For each broadcast domain, only one monitoring connection is needed.

\subsection{Validation of the Proposed Scheme}
To validate the method and the framework, a testbed assembled with the components of \autoref{tab_racksetup} is utilized.
By probing real hardware, the methodology introduced in Section \ref{sec:methodology} and the \ac{PoC} implementation are verified.
The used database comprises 8499 known devices excluding the devices from the testbed which are blacklisted for the validation.
The network traffic used to validate the accuracy is the same 12 hours pcap as that used for interarrival time calculation (see \autoref{fig_arpboxplot_big}).
\autoref{tab_validation} shows the results of the validation categorized into four possible results.

\textbf{Correct identification} (\cm):
As illustrated in Section \ref{sec:methodology_correlation}, a small distance to a known device indicates a correct identification.
A distance of 0x000000 indicates that the device was clearly identified. 
This happens if an already identified device should be identified again, e.g. in a previous scan or in a continuous monitoring process.

\textbf{Correct vendor, wrong device} (\tm): 
In a false positive identification, the identified device does not match the real one.
With an increasing distance between the MAC of the captured device and the entries in the database, the likelihood of a correct identification decreases.

\textbf{Only vendor identified} (\om): 
Furthermore, it is possible that there is no match, e.g. if there is no device in the MA-L address space, leading to a distance higher than 0xFFFFFF. 
It is mostly still feasible to identify the vendor looking up the \ac{MAC} vendor list(s).

\textbf{No identification} (\xm):
If there is no known device in the address space and no entry in the vendor lists, no identification is possible.

\begin{table}[H]
\caption{Validation Results}
\centering
\label{tab_validation}
\footnotesize
\begin{tabular}{l >{\ttfamily}c l l >{\ttfamily}c l}
\hline \hline
\textbf{Device} & \textbf{Distance}      & \textbf{Ident.}         & \textbf{Device}        & \textbf{Distance}      & \textbf{Ident.} \\         
\hline
00     & 0x19CE87      & \tm            & 12            & 0x19D31A      & \tm            \\  
01     & 0x089F19      & \cm            & 13            & 0x089F18      & \cm            \\  
02     & 0x0FE94D      & \tm            & 14            & 0x0FE76D      & \tm            \\  
03     & 0x00049C      & \cm            & 15            & 0x00075D      & \cm            \\  
04     & 0x000481      & \cm            & 16            & 0x000487      & \cm            \\  
05     & 0x117EDC      & \cm            & 17            & 0x117EA6      & \cm            \\  
06     & 0x00A214      & \cm            & 18            & 0x00A218      & \cm            \\  
07     & 0x001789      & \tm            & 19            & 0x00177F      & \tm            \\  
08     & 0x00A630      & \tm            & 20            & 0x00A614      & \tm            \\  
09     & 0x000102      & \cm            & 21            & 0x000235      & \cm            \\  
10     & 0x033E07      & \cm            & 22            & 0x0340DF      & \cm            \\  
11     & 0x0001AC      & \cm            & 23            & 0x00003D      & \cm            \\  
\hline
\multicolumn{3}{l}{\cm \hspace*{0.1cm} \scriptsize Correct identification}           & \multicolumn{3}{l}{\om \hspace*{0.1cm} \scriptsize Only vendor identified} \\  
\multicolumn{3}{l}{\tm \hspace*{0.1cm} \scriptsize Correct vendor, wrong device}     & \multicolumn{3}{l}{\xm \hspace*{0.1cm} \scriptsize No identification} \\
\end{tabular}
\normalsize
\end{table}

In total, the validation leads to a discovery rate of 100\% with a correct identification rate of 66.67\%,
demonstrating the feasibility of our method and software.
These results could be improved with a larger device database.

\subsection{Comparison with Other Tools}
\label{sec:comparison}
We used three active and three passive network scanner tools, and compared their results with our framework.
A popular active scanner used in the Censys project is ZGrab (Git version 6c81ce4)~\cite{durumeric2015search},
which supports the following \ac{SCADA} protocols: BACnet, DNP3, Niagara Fox, Modbus, and SimaticS7.
PLCScan (Git version 014480c)~\cite{plcscan} is one of the first active scanners for industrial networks.
Furthermore, the active scan of nmap (Version 7.60)~\cite{nmap} with \ac{NSE} scripts for BACnet,
Ethernet/IP, Modbus, Nigara Fox, Omron, PcWorx, ProConOS, and SimaticS7 is used.
For the passive scan, Netdiscover (Version 0.3), SinFp(Version 1.22), and p0f (Version 3.09b),
introduced in Section \ref{sec:relatedwork}, are used.

\begin{table*}[!t]
\caption{Comparison with Existing Tools}
\centering
\label{tab_comparision}
\resizebox{\textwidth}{!}{%
\begin{tabular}{l | l l l l l l l l l l l l l l l l l l l l l l l l | r}
\hline \hline
                   & \multicolumn{24}{c|}{Device}                                                                                          & Ident.\\
\textbf{Scanner}   & 00 & 01 & 02 & 03 & 04 & 05 & 06 & 07 & 08 & 09 & 10 & 11 & 12 & 13 & 14 & 15 & 16 & 17 & 18 & 19 & 20 & 21 & 22 & 23 & Rate \\
\hline
zgrab              & \cm& \xm& \xm& \xm& \xm& \xm& \xm& \xm& \xm& \cm& \cm& \xm& \cm& \xm& \xm& \xm& \xm& \xm& \xm& \xm& \xm& \cm& \cm& \xm& 25.00\% \\
PLCScan            & \xm& \xm& \xm& \xm& \cm& \xm& \xm& \xm& \xm& \cm& \cm& \xm& \xm& \xm& \xm& \xm& \cm& \xm& \xm& \xm& \xm& \cm& \cm& \xm& 25.00\% \\
Nmap               & \cm& \om& \cm& \om& \om& \om& \om& \cm& \cm& \cm& \om& \om& \cm& \om& \cm& \om& \om& \om& \om& \cm& \cm& \cm& \om& \om& 41.67\% \\
SinFp              & \xm& \xm& \xm& \xm& \xm& \xm& \xm& \xm& \xm& \xm& \xm& \xm& \xm& \xm& \xm& \xm& \xm& \xm& \xm& \xm& \xm& \xm& \xm& \xm& 00.00\% \\
p0f                & \xm& \xm& \xm& \xm& \xm& \xm& \xm& \xm& \xm& \xm& \xm& \xm& \xm& \xm& \xm& \xm& \xm& \xm& \xm& \xm& \xm& \xm& \xm& \xm& 00.00\% \\
Netdiscover        & \om& \om& \om& \om& \om& \om& \om& \om& \om& \om& \om& \om& \om& \om& \om& \om& \om& \om& \om& \om& \om& \om& \om& \om& 00.00\% \\
Our tool           & \tm& \cm& \tm& \cm& \cm& \cm& \cm& \tm& \tm& \cm& \cm& \cm& \tm& \cm& \tm& \cm& \cm& \cm& \cm& \tm& \tm& \cm& \cm& \cm& 66.67\% \\
\multicolumn{25}{r}{\cm \hspace*{0.1cm} Correct identification \hspace*{0.2cm} \om \hspace*{0.1cm} Only vendor identified \hspace*{0.2cm} \tm \hspace*{0.1cm} Correct vendor, wrong device \hspace*{0.2cm} \xm \hspace*{0.1cm} No identification} \\
\end{tabular}}
\end{table*}

To get comparable results, all tools are evaluated in the same testbed as introduced in Section \ref{sec:framework}.
The results of the comparison are shown in \autoref{tab_comparision}.
The identification rate of the active scanners depends on the implementation of the protocol used in \ac{SCADA} environments.
If the protocol is available for the scanner, then it mostly discovers and identifies the device.  
ZGrab and PLCScan have found fewer \acp{PLC}, because there are fewer protocols implemented than Nmap with additional \ac{NSE} scripts.
Similar to our tool, Netdiscover uses the IEEE OUI vendor list and was able to detect the vendors of the devices but not the specific product.
SinFp and p0f could not identify any \ac{ICS} device correctly in our testbed.
They only showed office components such as the server used for the evaluation.
Similar results of the existing fingerprinting tools were achieved by others in the past~\cite{Caselli2013,hahn2011evaluation}.
Our framework was able to discover all devices in the testbed by their vendor, and about 66\% of the products were correctly identified.

\subsection{Discussion of Limitations}
\label{sec:restrictions}
The proposed \ac{MAC}-based device identification has some limitations regarding the rate of discovery and identification. Some of these limitations, however, do not apply to industrial networks due to their structure and specific use of the connected devices.

\subsubsection{MAC Address Spoofing}
IT devices often allow specifying a MAC address of a network interface.
For example, a network administrator wants to clone a device. 
Nevertheless, this is mostly not possible for \ac{ICS} devices, because vendors do not provide this feature. 
Of course, attackers could bring in new devices into a network by sending \ac{ARP} requests with spoofed \ac{MAC} addresses.
However, attackers cannot spoof \ac{ARP} requests of existing devices without physical access or access to the switching hardware. 
Therefore, \ac{MAC} address spoofing has only minor relevance for the proposed scheme.

\subsubsection{Database Quality}
To identify devices by their \ac{MAC} address, there must be a similarly known device in the local database. 
Hence, the number of entries in the known \ac{MAC} address database is significant for the quality of device identification.
This could be improved by community contributions.

\subsubsection{MAC Randomization for Connections}
In order to make it difficult to identify devices, it is possible to randomly choose \ac{MAC} addresses for every new connection.
This could lead to a false positive identification of a device.
However, for stationary devices, especially in industrial networks like \acp{PLC}, it is uncommon to randomize the \ac{MAC} address.

\subsubsection{Vendor Assignment Process}
Vendors sometimes assign the same \ac{MAC} address to multiple devices, maybe by mistake or intentionally to save money.
This is not a major problem if the devices are shipped to different parts of the world.
In the case of \ac{MAC}-based identification, this could lead to false identifications if different products have the same \ac{MAC}.

\subsubsection{Static \ac{ARP} Table}
In a static \ac{ARP} table, the \ac{MAC} addresses of some or all network participants have fixed entries. 
It is used in static environments, e.g. to prevent \ac{MitM} attacks by \ac{ARP} poisoning, to use a network diode, or to reduce the broadcast traffic~\cite{stouffer2011guide}. 
In this case, the \ac{ARP} broadcasts used to identify the devices will not be sent, and thus, no identification is possible.

\subsubsection{Firmware/Software Version}
The firmware and software versions are not detected by the \ac{MAC}-based identification of a device. 
Consequently, the vulnerability allocation is done without considering the software version. 
However, in many \acp{ICS}, the devices are still in the delivery software state and are not patched. 
This information could be used to roughly estimate the firmware version of the device, which is sufficient for an initial assessment.

\section{Conclusion}
\label{sec:conclusion}
This paper has introduced a new method for passive scanning of critical infrastructures and provided a functional \ac{PoC}. 
With a \ac{MAC}-based discovery and identification of \ac{ICS} components, a safe passive scan within \acp{ICS} is possible.
We have shown the feasibility of the method with a validation resulting in a total discovery rate of 100\% and an identification quote of more than 66\%.
In comparison with existing tools, the \ac{MAC} address correlation approach performed well. 
The minimal integration effort and the extensibility of our framework have advantages over classical scan methods and deep packet inspection.

Further, we plan to publish the code of our framework and are open for contribution to the databases. 
The material includes the source code of the framework, network captures, and additional information about the testbed.

With a detailed recognition of the transitions between products during the \ac{MAC} address assignment, an even more detailed identification could be achieved.
Therefore, some details from the vendors or a larger database of each vendor are necessary due to the vendor-specific assignment process.
Additionally, the users of our approach can add known devices from an industrial plant with which similar devices in other plants could be identified.

\section*{Acknowledgement}
The work on MAC-based device discovery and identification is part of the RiskViz~\cite{Riskviz} research project, funded by the German \ac{BMBF}. 
We would also like to thank Georg Sigl (Fraunhofer AISEC) for his valuable feedback.

\bibliography{\jobname}

\begin{acronym}
 \acro{A}{Availability}
 \acro{ACK}{acknowledgment}
 \acro{ARP}{Address Resolution Protocol}
 \acro{BMBF}{Federal Ministry of Education and Research}
 \acro{C}{Confidentiality}
 \acro{CIA}{Confidentiality, Integrity and Availability}
 \acro{CID}{Company IDentifier}
 \acro{CPE}{Common Platform Enumeration}
 \acro{CSV}{Comma-Separated Values}
 \acro{CVE}{Common Vulnerabilities and Exposures}
 \acro{DCS}{Distributed Control System}
 \acrodefplural{DCS}{Distributed Control Systems}
 \acro{DHCP}{Dynamic Host Configuration Protocol}
 \acro{DoS}{Denial of Service}
 \acro{DuT}{Device under Test}
 \acro{ERP}{Enterprise Resource Planning}
 \acro{HMI}{Human Machine Interface}
 \acrodefplural{HMI}{Human Media Interfaces}
 \acro{I}{Integrity}
 \acro{IEEE}{Institute of Electrical and Electronics Engineers}
 \acro{ICS}{Industrial Control System}
 \acrodefplural{ICS}{Industrial Control Systems}
 \acro{IDE}{Integrated Development Environment}
 \acro{IoT}{Internet of Things}
 \acro{IIoT}{Industrial Internet of Things}
 \acro{IP}{Internet Protocol}
 \acro{MAC}{Media Access Control}
 \acro{MES}{Manufacturing Execution System}
 \acro{MitM}{Man-in-the-Middle}
 \acro{NSE}{Nmap Scripting Engine}
 \acro{OS}{Operating System}
 \acro{OSI}{Open Systems Interconnection}
 \acro{OT}{Operational Technology}
 \acro{OUI}{Organizationally Unique Identifier}
 \acro{pcap}{packet capture}
 \acrodefplural{pcap}{packet captures}
 \acro{PLC}{Programmable Logic Controller}
 \acrodefplural{PLC}{Programmable Logic Controllers}
 \acro{PoC}{Proof of Concept}
 \acro{RA}{Registration Authority}
 \acro{SCADA}{Supervisory Control and Data Acquisition}
 \acro{SYN}{synchronize}
 \acro{TAP}{Terminal Access Point}
 \acro{TCP}{Transmission Control Protocol}
 \acro{ToS}{Type of Service}
 \acro{TTL}{Time to Live}
 \acro{USB}{Universal Serial Bus}
 \acro{VLAN}{Virtual Local Area Network}
 \acrodefplural{VLAN}{Virtual Local Area Networks}
 \acro{VPN}{Virtual Private Network}
\end{acronym}

\end{document}